%% file: main.tex
\documentclass{ieeeaccess}

\usepackage{cite}
\usepackage{amssymb}
\usepackage{graphicx}
\usepackage{textcomp}

\usepackage[utf8]{inputenc} 
\usepackage[T1]{fontenc}    
\usepackage{hyperref}       
\usepackage{url}            
\usepackage{booktabs}       
\usepackage{amsfonts}       
\usepackage{nicefrac}       
\usepackage{microtype}      
\usepackage{algorithm} 
\usepackage{algpseudocode}
\usepackage{mathtools}
\usepackage{amsmath}
\usepackage{upgreek}
\usepackage{multicol}
\usepackage{subfig}
\usepackage{graphicx}
\usepackage{bookmark}
\usepackage{multirow}
\usepackage{tabularx,booktabs}

\newcolumntype{C}{>{\centering\arraybackslash}X} 
\DeclareMathOperator*{\argmax}{argmax} 

\def\BibTeX{{\rm B\kern-.05em{\sc i\kern-.025em b}\kern-.08em
    T\kern-.1667em\lower.7ex\hbox{E}\kern-.125emX}}
    
\begin{document}

\history{Received 10 April 2023, accepted 5 May 2023, date of publication 12 May 2023, date of current version 18 May 2023.}
\doi{10.1109/ACCESS.2023.3275883}

\title{Multi-Agent Reinforcement Learning Based on Representational Communication for Large-Scale Traffic Signal Control}

\author{
    \uppercase{Rohit Bokade} \authorrefmark{1}, 
    \uppercase{Xiaoning Jin} \authorrefmark{1},  ~\IEEEmembership{Member, IEEE},
    and ~ \uppercase{Christopher Amato} \authorrefmark{2} 
}
\address[1]{Department of Mechanical and Industrial Engineering, Northeastern University, Boston, MA 02115, USA}
\address[2]{Khoury College of Computer Sciences, Northeastern University, Boston, MA 02115, USA}

\markboth
{Bokade \headeretal: MARL Based on Represntational Communication for Large-Scale TSC}
{Bokade \headeretal: MARL Based on Represntational Communication for Large-Scale TSC}

\input{abstract}
\input{keywords}
\titlepgskip=-15pt
\maketitle
\input{introduction}

\input{related_work}

\input{background}

\input{proposed_framework}

\input{experiments}

\input{conclusion}
\input{acknowledgement}
\bibliography{references}
\bibliographystyle{ieeetr}

\begin{IEEEbiography}[{\includegraphics[width=1in,height=1.25in,clip,keepaspectratio]{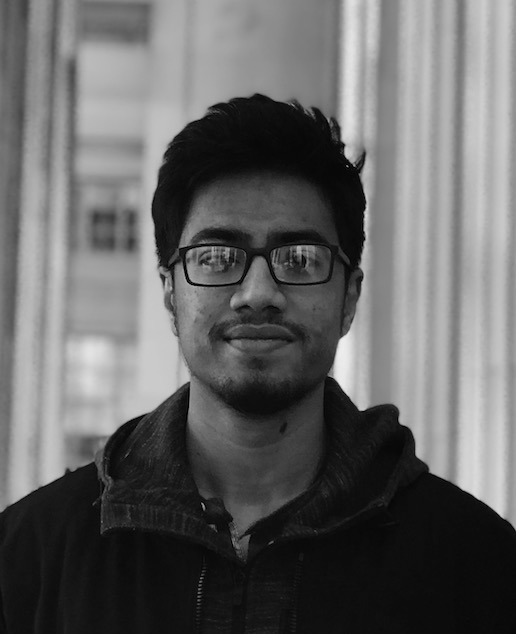}}]{Rohit Bokade}
    received his Masters degree in Operations Research from Northeastern University, where he is currently pursuing a Ph.D. degree in Industrial Engineering. His current research interests involve exploring the potential of advanced machine learning techniques, such as reinforcement learning, deep learning, and optimization techniques to improve industrial engineering practices and solve real-world problems.
\end{IEEEbiography}

\begin{IEEEbiography}[{\includegraphics[width=1in,height=1.25in,clip,keepaspectratio]{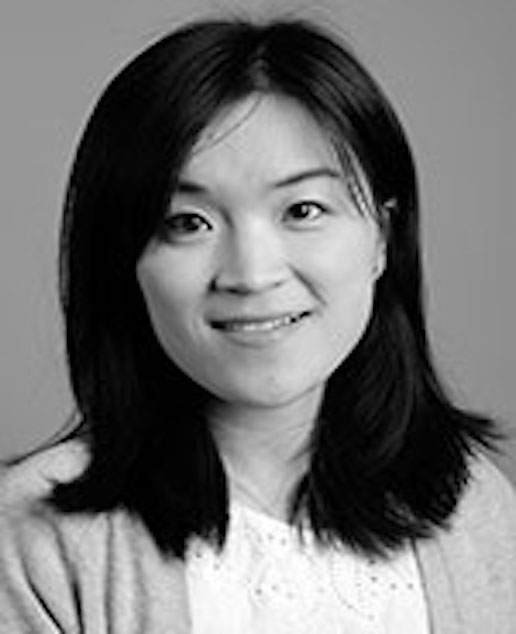}}]{Xiaoning Jin}
    (Member, IEEE) received the Ph.D. degree in Industrial and Systems Engineering from the University of Michigan, Ann Arbor, MI, USA in 2012.
    She is currently an Assistant Professor of mechanical and industrial engineering with the College of Engineering at Northeastern University, Boston, USA.  She is the recipient of the National Science Foundation Career Award in 2020. She has over 50 papers in fully refereed international journals and conferences. Her research interests include predictive analytics and decision making, data analytics, fault diagnostics and prognostics, and artificial intelligence in various engineering applications.
    Prof. Jin currently serves as the Vice-Chair of the Manufacturing Systems Technical Committee with the ASME Manufacturing Science and Engineering division. She received the 2016 Outstanding Young Manufacturing Engineer Award from the Society of Manufacturing Engineers (SME).
\end{IEEEbiography}

\begin{IEEEbiography}[{\includegraphics[width=1in,height=1.25in,clip,keepaspectratio]{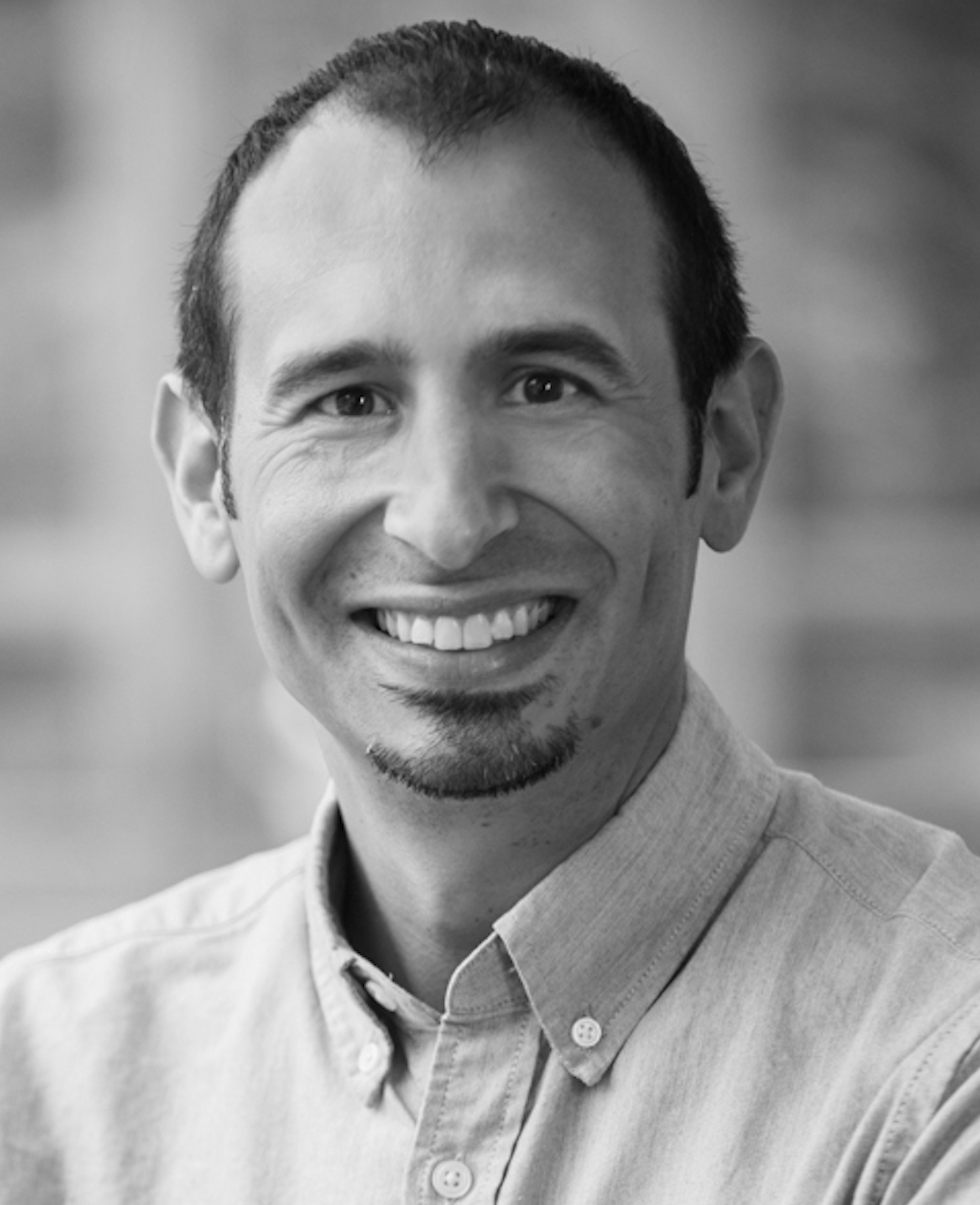}}]{Christopher Amato}
    is an Assistant Professor at Northeastern University where he leads the Lab for Learning and Planning in Robotics. Before joining Northeastern, Dr. Amato was a Research Scientist at Aptima, Inc. and a Postdoc and Research Scientist at MIT as well as an Assistant Professor at the University of New Hampshire. He has published many papers in leading artificial intelligence, machine learning and robotics conferences (including winning a best paper prize at AAMAS-14 and being nominated for the best paper at RSS-15, AAAI-19, AAMAS-21 and MRS-21). He has also won several awards such as Amazon Research Awards and an NSF CAREER Award. His research focuses on reinforcement learning and planning in partially observable and multi-agent/multi-robot systems.
\end{IEEEbiography}

\newpage
\onecolumn
\appendix
\input{appendix}

\EOD
\end{document}

%% file: abstract.tex
\begin{abstract}

Traffic signal control (TSC) is a challenging problem within intelligent transportation systems and has been tackled using multi-agent reinforcement learning (MARL). While centralized approaches are often infeasible for large-scale TSC problems, decentralized approaches provide scalability but introduce new challenges, such as partial observability. Communication plays a critical role in decentralized MARL, as agents must learn to exchange information using messages to better understand the system and achieve effective coordination. Deep MARL has been used to enable inter-agent communication by learning communication protocols in a differentiable manner. However, many deep MARL communication frameworks proposed for TSC allow agents to communicate with all other agents at all times, which can add to the existing noise in the system and degrade overall performance. In this study, we propose a communication-based MARL framework for large-scale TSC. Our framework allows each agent to learn a communication policy that dictates "which" part of the message is sent "to whom". In essence, our framework enables agents to selectively choose the recipients of their messages and exchange variable length messages with them. This results in a decentralized and flexible communication mechanism in which agents can effectively use the communication channel only when necessary. We designed two networks, a synthetic $4 \times 4$ grid network and a real-world network based on the Pasubio neighborhood in Bologna. Our framework achieved the lowest network congestion compared to related methods, with agents utilizing $\sim 47-65 \%$ of the communication channel. Ablation studies further demonstrated the effectiveness of the communication policies learned within our framework.

\end{abstract}

%% file: keywords.tex

%% file: introduction.tex
\section{Introduction}

Rapid urbanization in recent years \cite{2014RevisionWorld2014} has given rise to a growing problem of traffic congestion \cite{colakUnderstandingCongestedTravel2016}. Recent trends also show a huge rise in ride-hailing and e-commerce services, which have contributed significantly towards the increasing number of vehicles on the road \cite{us2016stats, schaller2018new}. The impacts of traffic congestion include increased delays and wasted fuel in addition to the impact on the environment and public health \cite{schrank2009tti,levy2010evaluation}. Traffic signal control (TSC) is one of the challenging bottlenecks in reducing traffic congestion. The goal of TSC is to dynamically and intelligently control signal timings to reduce the number of vehicles halted on the road.

Recent advances in machine learning have opened up a wide range of opportunities for developing intelligent transportation systems solutions, including traffic signal control. Deep learning based architectures provide flexibility in processing data from various sensory inputs \cite{alomStateoftheArtSurveyDeep2019} and additionally serve as a useful tool for multimodal data fusion \cite{bokade2021cross}. Deep reinforcement learning (RL) uses deep neural networks (DNNs) to map inputs to actions. Deep RL frameworks have shown tremendous progress in learning effective policies directly from raw sensory inputs \cite{mnih2013playing}. Following these advances, deep MARL has emerged as one of the promising tools to develop effective frameworks for network-wide TSC, where each traffic light is treated as an agent that learns to select appropriate phases to minimize congestion within the network.
 
A straightforward way to carry over the framework of deep RL into the MARL setting is to treat all the agents as a collective entity. One can then use a function approximator, such as DNNs, to map the state into joint actions. However, the problem with this approach is that the action space grows exponentially with the number of agents. This kind of centralized control often proves impractical for large-scale applications. Furthermore, centralized approaches require access to the global state of the environment, which may not always be feasible. TSC is a large-scale problem for which decentralized execution becomes crucial. Several deep MARL frameworks have been proposed for independently controlling the traffic signals 
\cite{kuyer2008multiagent,zhao2011computational,genders2016using,van2016coordinated,calvo2018heterogeneous,liang2018deep,lin2018efficient,camelo2019parallel,chu2019multi,tan2019cooperative,wei2019survey,wei2019colight,wei2019presslight,zheng2019learning,gupta2020networked,haydari2020deep,jaleel2020reducing,ma2020feudal,tan2020multi,wang2020stmarl,wu2020edge,xie2020iedqn,zang2020metalight,zhao2020learning,devailly2021ig,liu2021learning,wang2021traffic,zhu2021variationally}. However, to apply these methods to real-world applications, such as TSC, one must consider potential limitations of communication such as bandwidth availability \cite{pynadath2002communicative}. In addition, allowing such unrestricted communication can be disadvantageous for several reasons. One reason is that the system incurs additional communication overhead when the messages received by an agent are unhelpful and excess communication can increase the overhead and reduce performance by adding unnecessary noise. Another reason is that it leaves the system in a state of vulnerability to adversarial attacks. Potential solutions to these problems are (1) compressing the information into a small number of bits \cite{foerster2016learning,sukhbaatar2016learning,hoshen2017vain,pesce2020improving}, (2) communicating only when necessary \cite{zhang2019efficient,singh2018learning,wang2019learning,kim2019learning,niu2021multi,agarwal2019learning,du2021learning}, and/or (3) communicating with selective agents \cite{jiang2018graph,liu2020multi,zhang2019efficient,wang2019learning,agarwal2019learning,du2021learning}. The majority of studies that proposed message passing mechanisms focused extensively on the aspect of improving the content of the messages by leveraging techniques from DL (e.g., attention mechanism \cite{vaswani2017attention}, graph neural networks \cite{scarselli2008graph}, and variational inference \cite{alemi2016deep}). The lines of work that focused on addressing the problem of deciding \textit{when to communicate} or \textit{whom to communicate with} involved heuristics-based frameworks \cite{zhang2019efficient,wang2019learning}, or use gating mechanisms \cite{jiang2018graph,singh2018learning,kim2019learning,liu2020multi,du2021learning}.

\subsection{Contribution}

In this paper, we propose an alternate framework for learning communication protocols that builds upon the existing Q-MIX \cite{rashid2020monotonic} and NDQ \cite{wang2019learning} frameworks, which leverage the paradigm of centralized training and decentralized execution (CTDE) by learning a global action-value function. The global action-value is monotonically decomposed into individual action-values for decentralized execution. Facilitating communication among agents results in better action-value estimates \cite{wang2019learning}. Within our framework, QRC-TSC, agents learn how to effectively compress their environmental perception and action intentions into a message and determine which part of the message needs to be transmitted to another agent for effective coordination. Also, this decision is made independently for each available recipient, thereby, making the communication framework flexible. We utilize the \textit{variational inference deep learning} framework \cite{alemi2016deep,jang2016categorical,maddison2016concrete,dupont2018learning} to maximize the mutual information between the message sent by the sender and the actions taken by the recipient \cite{wang2019learning}, which is an effective metric to measure communication performance \cite{lowe2019pitfalls}. Specifically, we model the message space as a joint distribution of generated message and communication policy (whether to send the bit of message). Through our formulation of the communication objective, we also encourage exploration over the communication policy space.

\begin{figure}[!ht]
    \centering
    \includegraphics[width=0.4\textwidth]{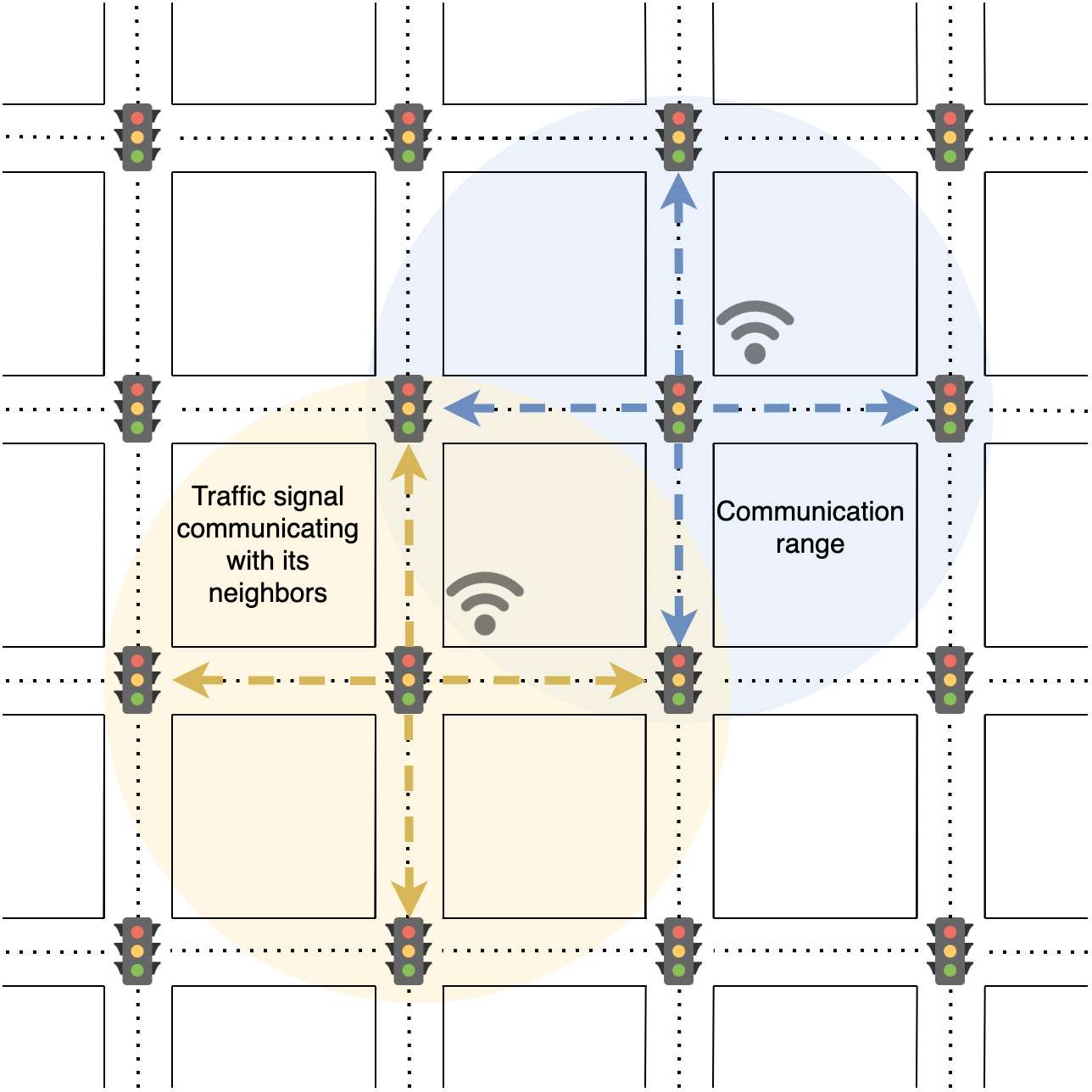}
    \caption{The highlighted circles represent communication range for each traffic light, i.e., each traffic light can communicate with its immediate neighbor or within 500 meters of range.}
\end{figure}

We used the SUMO simulator \cite{krajzewicz2012recent} to design two traffic networks, a $4 \times 4$ synthetic grid network with variable traffic flow and a real-world network based on the city of Pasubio. We demonstrated the efficacy of our framework in reducing the congestion level of network-wide traffic by comparing it with some of the leading communication-based MARL frameworks. We also conducted ablation studies on the communication mechanism by comparing the results of our framework with several baseline communication strategies, including full communication, no communication, and random communication. We observed that traffic signals on the network were able to dynamically adjust the number of bits they send in the messages while maximizing performance.

The rest of the paper is organized as follows. Section \ref{sec:background} provides an overview of the relevant work done in MARL which serves as the basis of our framework. Section \ref{sec:proposed_framework} discusses our framework in detail and also describes the formulation of the TSC problem within our framework. In Section \ref{sec:experiments} we provide the experimental setup and compare the results of QRC-TSC with other frameworks and perform ablation studies. Finally, Section \ref{sec:conclusion} concludes the paper and discusses potential future research directions.

%% file: related_work.tex
\section{Related work}
\label{sec:related_work}

\subsection{Inter-agent communication in MARL}

Recently proposed algorithms (e.g., DIAL \cite{foerster2016learning} and CommNet \cite{sukhbaatar2016learning}) have made it possible to learn communication protocols through a feedback mechanism by leveraging DL techniques. DIAL is an extension to Independent Q-learning (IQL), where each agent generates both action-values and a message vector. The message vector is then passed as input to the other agent networks in the next time step, thus obtaining feedback from the receiver agents in the form of gradients.

The most relevant work to our problem is the Nearly Decomposable Q-function (NDQ) \cite{wang2019learning}, which combines the communication framework of DIAL with the general learning framework of Q-MIX by utilizing the variational inference \cite{alemi2016deep} technique from DL. In addition to learning communication through feedback, NDQ proposes an objective function that maximizes the mutual information (MI) between the sender's message and the recipient's action. The main idea is for agents to learn to capture the most relevant information in as few bits of message as possible. A similar metric, causal influence of communication (CIC), was proposed \cite{lowe2019pitfalls,jaques2019social} to improve communication performance without impeding the general learning process. However, NDQ uses a threshold-based heuristic to filter out unhelpful messages in its communication framework. In our work, we extend the work done in NDQ and develop a communication framework that learns to effectively select the important bits of messages.

\subsection{Deep Multi-Agent Reinforcement Learning in Traffic Signal Control}

The problem of TSC has been studied through the lens of MARL \cite{wiering2000multi,wiering2004intelligent,kuyer2008multiagent} by treating the traffic signal as an agent and rewarding it based on a metric that is inversely proportional to the level of congestion (queue length). Recently, with the advent of Deep MARL, many proposed solutions to the problem of TSC \cite{genders2016using,van2016coordinated,chu2019multi,tan2019cooperative,wei2019colight,tan2020multi,wang2020large,zang2020metalight,zhu2021variationally} were effective in extracting richer information from more sophisticated sensor inputs for the decision-making process \cite{wei2019survey,haydari2020deep}. Communication mechanisms are a part of the progress in applying MARL in TSC domains as well. Several methods proposed for TSC \cite{el2013multiagent,wei2019survey,chu2020multi,ma2020feudal,wang2020large} implemented a variety of communication mechanisms to train the traffic signals to send and receive messages from neighboring traffic signals. However, the aforementioned methods fail to avoid the pitfall of unrestricted communication. TSC is a large-scale problem where communication between traffic signals has to be wireless, which comes at the cost of limited bandwidth and requires the utilization of additional resources. Hence, the communication mechanism must be efficient in allowing traffic signals to exchange relevant information only when it is beneficial. \\

%% file: background.tex
\section{Background}
\label{sec:background}

\subsubsection{Deep reinforcement learning}

Reinforcement learning (RL) aims at learning the optimal policy through repeated interaction with the environment. A standard RL problem can be formulated as a Markov Decision Process (MDP). At each time step $t$ agent observes the state of the environment $s_{t} \in S$ and takes an action $a_{t} \in A$ according to policy $\pi$. Based on this action, the agent receives feedback from the environment in the form of reward $r_{t}$ and transitions to the next state $s_{t+1}$. The objective is to maximize the total expected discounted reward $R = \sum_{t=1}^{T} \gamma^{t} r_{t}$, where $\gamma \in [0, 1]$ is the discount factor.  

Deep Q-Networks (DQN) learns the action-value function $$Q_\theta = E[R_{t} | s_{t}=s, a_{t}=a],$$ where $\theta$ represents the parameters of the Q-network. The action-value function can be trained recursively by minimizing the loss $$\mathcal{L}(\theta) = E_{s, a, r, s'}[(y - Q_{\theta}(s, a))^{2}],$$ where $y = r + \gamma \max_{a'}Q_{\theta'}(s', a')$ and $\theta '$ represents the parameters of the target network. The agent selects the action that maximizes the Q-value with the probability $1 - \epsilon$ or acts randomly with probability $\epsilon$. The set of parameters $\theta^{-}$ in the target network are updated in regular time intervals by copying over the parameters $\theta$ from the primary network. Double DQN \cite{van2016deep} modifies DQN to add stabilization and avoid overestimation. In Double DQN, the target action-value is indexed from the output of the target network based on the greedy action selected by the primary network $$ y = r + Q_{\theta'}(\argmax_{a'}Q_{\theta}(\cdot | s') | s').$$ Both DQN and Double DQN are based on fully observable MDPs. However, in partially observable settings, an agent conditions its action-value function on the action-observation history. DRQN \cite{hausknecht2015deep} achieves this by using recurrent neural networks. At each time step, the Q-network takes as input the observation $o_{t}$, and the hidden state $h_{t - 1}$ to approximate the action values $Q_{\theta}(o_{t}, h_{t - 1}, a_{t})$. This enables the agent to integrate past information to make decisions.

\subsection{Cooperative Deep Multi-Agent Deep Reinforcement Learning}

One approach to modeling multi-agent systems as RL problems is to treat the whole system as a single agent. The agent observes the true state of the environment and selects joint-actions for all the agents. This approach, however, scales poorly as the search space for joint-action increases exponentially with the number of agents in the system. A more feasible approach is to enable each agent to act independently. Thus, one could formulate the problem as a decentralized partially observable Markov decision process (Dec-POMDP), which extends the framework of MDP to multi-agent scenarios with partial observability \cite{oliehoek2016concise}. It is defined by a tuple of $\mathcal{M} = <\mathcal{S}, \mathcal{A}, \mathcal{P}, \mathit{\Omega}, \mathcal{O}, \mathit{r}, \mathcal{N}, \mathit{\gamma}>$, where $\mathit{s} \in \mathcal{S}$ is the global state space and $\mathit{i} \in \mathcal{N} \equiv \{ 1, \cdots, n \}$ is the finite set of agents. At time step $t$, each agent $a$ selects an action $a^{i} \in \mathcal{A}$ resulting in a joint action vector $a \in \mathbf{\mathcal{A}} \equiv \mathcal{A}^{n}$. The transition dynamics of the environment state are given by $\mathit{P}(s'|s, a)$. All agents receive a shared reward according to the reward function $r(s, a)$ and $\gamma \in [0, 1)$ is the discount factor. Each agent receives an observation $o^{a} \in \Omega$ according to the observation function $O(s, a)$. Each agent has an action-observation history $\tau^{i} \in \mathcal{T} \equiv (\mathit{\Omega} \times \mathcal{A})^{*}$ on which it conditions its individual policy $\pi^{i}(a^{i}|\tau^{i})$. The joint policy $\boldsymbol{\pi} = <\pi^{1}, \cdots, \pi^{n}>$ induces a joint-action value function $$Q^{\pi}(s, \mathbf{a}) = E_{s_{0:\infty};\mathbf{a}_{0:\infty}}[\sum_{t=0}^{\infty} \gamma^{t}r_{t}|s_{0} = s; \mathbf{a}_{0} = \mathbf{a}, \boldsymbol{\pi}].$$

Some studies propose that each agent learn the global action-value \cite{tan1993multi}. Recent works have demonstrated better performance with monotonic factorization of the global-action value \cite{sunehag2017value,rashid2020monotonic}. Q-MIX \cite{rashid2020monotonic}, specifically, leverages the CTDE paradigm to learn a monotonic mapping between individual utilities and the global action-value by utilizing a mixing network $Q_{total}(\tau, a) = f(Q_{1}(\tau^{1}, a^{1}), \cdots, Q_{n}(\tau^{n}, a^{n}); \theta_{mixer}).$ The weights of the mixing network $\theta_{mixer}$ are generated by a set of hypernetworks, conditioned on the state $s_{t}$, with absolute activation function to ensure monotonicity $\frac{\partial Q_{total}}{\partial Q_{i}} \geq 0$. The decomposition allows for decentralized action selection during execution, since the mixing network is only used for training. Thus, the mixing network can be conditioned on additional information available during the training. Recent works improved performance on complex multi-agent environments by combining Q-MIX with communication framework \cite{wang2019learning,zhang2019efficient}. Thus, we utilize Q-MIX as the base framework for our proposed communication mechanism.

\begin{align*}
    & \argmax_{a}Q_{total}(\tau, \mathbf{a}) = \\
    & \left (\argmax_{a^{1}}Q_{1}(\tau^{1}, a^{1}), \cdots, \argmax_{a^{n}}Q_{n}(\tau^{n}, a^{n}) \right) \\
\end{align*}

%% file: proposed_framework.tex
\section{Proposed Framework}
\label{sec:proposed_framework}

\subsection{Problem Formulation}

We extend the framework of Dec-POMDP to incorporate inter-agent communication. We formulate the traffic signal network as an undirected graph $\mathcal{G} = (\mathcal{V}, \mathcal{E})$, where $v_{i} \in \mathcal{V}$ is the set of nodes and $v_{ij} \in \mathcal{E}$ is the set of edges. Each node represents an agent (traffic signal) and each edge represents the connectivity between agents. The neighborhood for a node $v$ is defined as $\mathcal{N}(v) =\{  u \in \mathcal{V} | (u, v) \} \in \mathcal{E}$ and the adjacency matrix $\mathbf{A}$ is a $n \times n$ matrix with $A_{ij} = 1$ if $e_{ij} \in \mathcal{E}$ and $A_{ij} = 0$ if $e_{ij} \notin \mathcal{E}$. We design communication framework such that each agent is only allowed to communicate with its neighbors.

We set up the problem of TSC as a Dec-POMDP, where each traffic signal in the network is treated as an agent and the central goal of the system is to reduce network-wide congestion. The traffic signals make decisions using information about incoming vehicles, which is assumed to be accessible through sensors located near the signals. The traffic signals control the flow of traffic through the intersection by selecting a phase from the available set of phases. We discuss the details of our formulation in detail below.

\begin{figure}[!htb]
    \centering
    \includegraphics[width=0.3\textwidth, height=8.5cm]{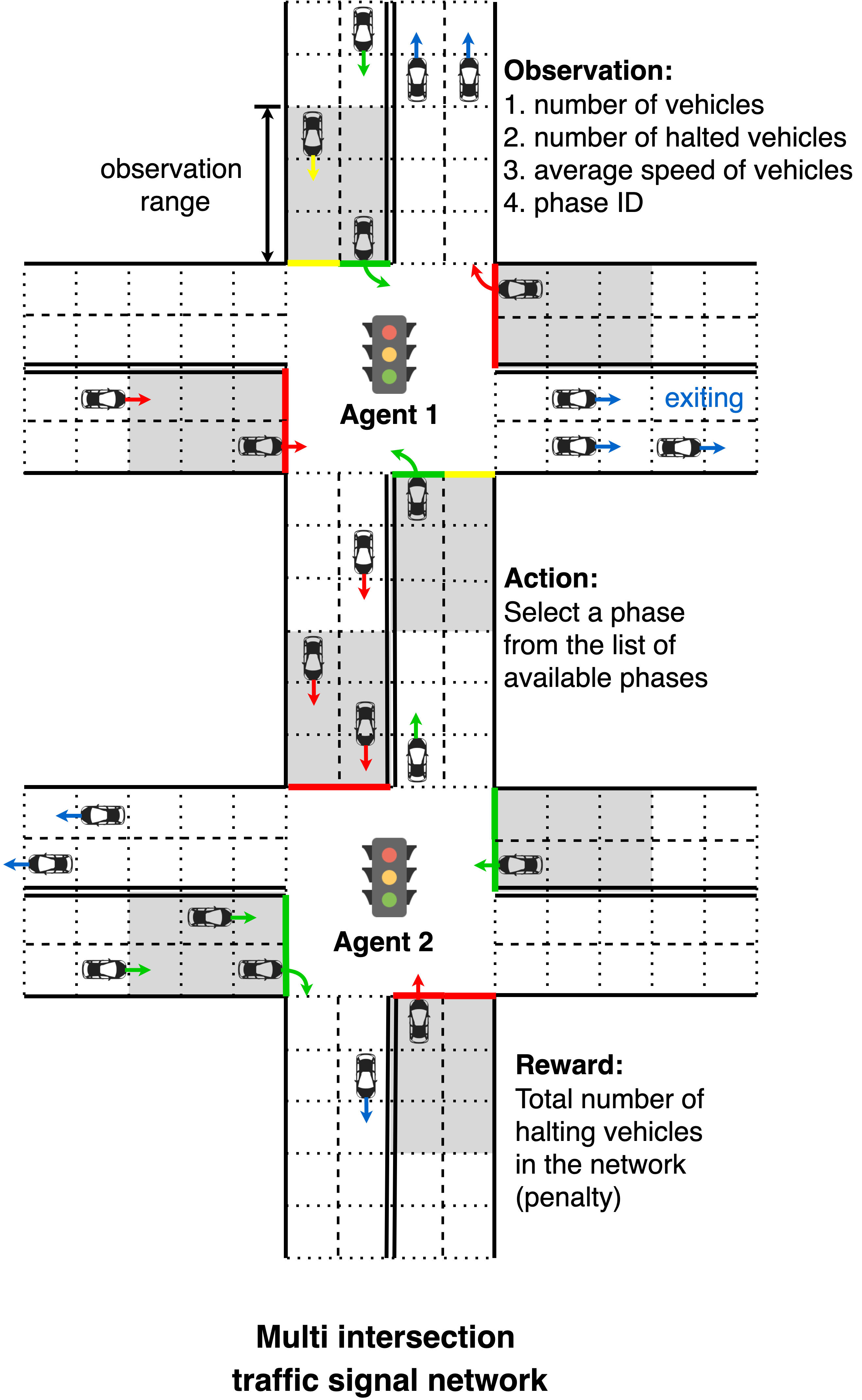}
    \caption{Prototype of a traffic signal network with two intersections. The highlighted zones on the incoming lane on each traffic light represent the range within which the traffic light can access information about the vehicles.}
    \label{fig:traffic_signal_observations}
\end{figure}

\subsubsection{Observation representation}

Each traffic signal has a limited range of vision of 50 meters, within which it can obtain information related to the traffic flow. This is equivalent to the sensory information that can be obtained from practical common sensors. We implement observation collection in the environment using by placing \texttt{laneAreaDetector} of length 50 meters on each incoming lane to capture the traffic information which can be seen by the boxes highlighted in grey in Fig. \ref{fig:traffic_signal_observations}. The observation for each traffic signal consists of: the number of vehicles $\{ n_{l} \}_{l=1}^{L_{i}}$, the average normalized speed of the vehicles $\{ s_{l} \}_{l=1}^{L_{i}}$, the number of halted vehicles (queue lengths) $\{ q_{l} \}_{l=1}^{L_{i}}$, and the current \texttt{phaseID} of the traffic signal, where $L_{i} \in L$ are the incoming lanes for a traffic signal $i$ and $L$ is a set of all the lanes in the network.

\subsubsection{Action Representation}

\begin{figure}[!htb]
    \centering
    \includegraphics[width=0.5\textwidth]{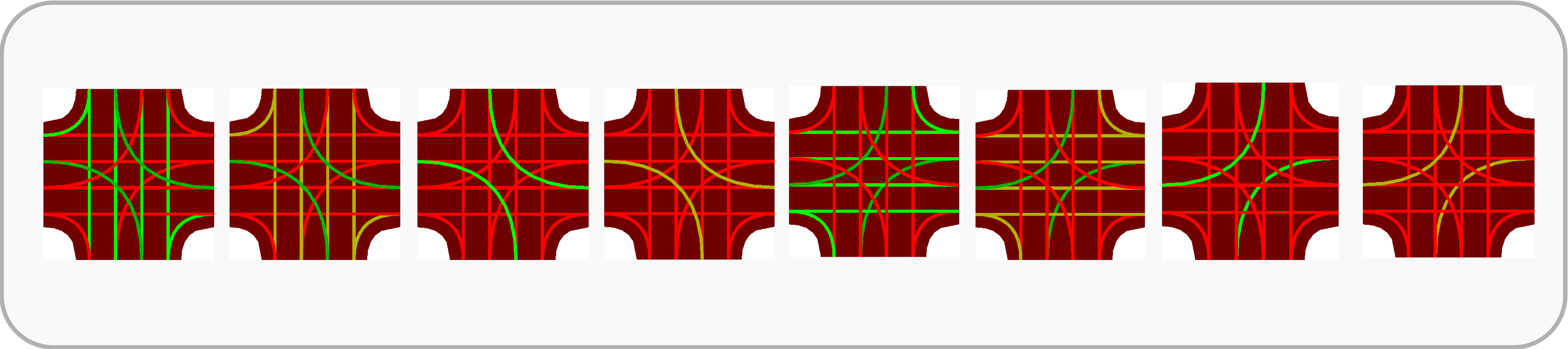}
    \caption{An example of the phases available for an intersection in a $4 \times 4$ grid network from SUMO simulator. The colored lines (red, yellow, and green) together indicate the phase of the traffic signal. The first phase (from the left) indicates an all green phase, where the vehicles are allowed to go straight and/or make turns. Each agent controls the traffic signals by selecting one of these phases.}
    \label{fig:4x4_intersection_phases}
\end{figure}

For each traffic signal $i$, we define its action $a_{i}$ as choosing one green phase from a list of available phases. As an example, Fig. \ref{fig:4x4_intersection_phases} shows the list of phases that are available for a traffic signal in a $4 \times 4$ grid network. A traffic signal can select any green phase from its list or keep its current one, but it must then follow the next yellow phase, which is enforced by the environment. The action selection interval and the yellow phases are fixed for a duration of 5 simulation seconds.

\subsubsection{Reward}

Various metrics are used for rewards in traffic signal control settings. In our study, we chose queue length $q_{l}$ as the performance metric of the traffic signal controller due to its simplistic nature and its property of representing an instantaneous feedback signal. We define the objective function as minimizing the number of vehicles stopped throughout the network $$ $$ where $r_{t} \in \mathbb{R}$ is the global reward and $l \in L$ represents the lanes in the network.

\subsection{Overall Framework}

\begin{figure}[!ht]
    \centering
    \includegraphics[width=0.45\textwidth]{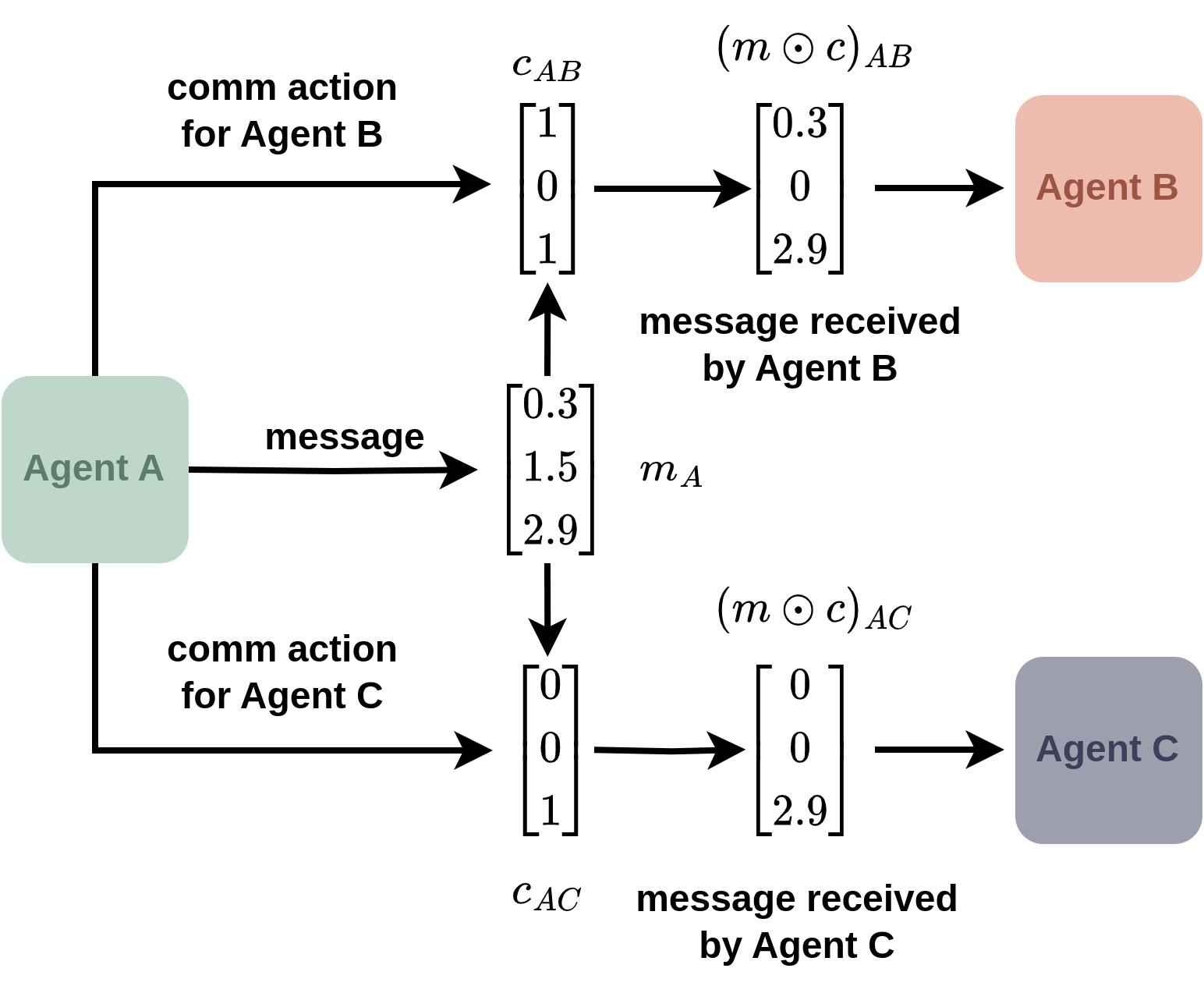}
    \caption{Example of the proposed communication framework. Agent A generates a message space $m_{A}$ and communication action $c_{AB}$ and $c_{AC}$ for agents B and agent C respectively. The message is then gated based on the communication action and sent to respective agents.}
    \label{fig:communication_concept}
\end{figure}

\begin{figure*}[!htb]
    \centering
    \includegraphics[width=\textwidth]{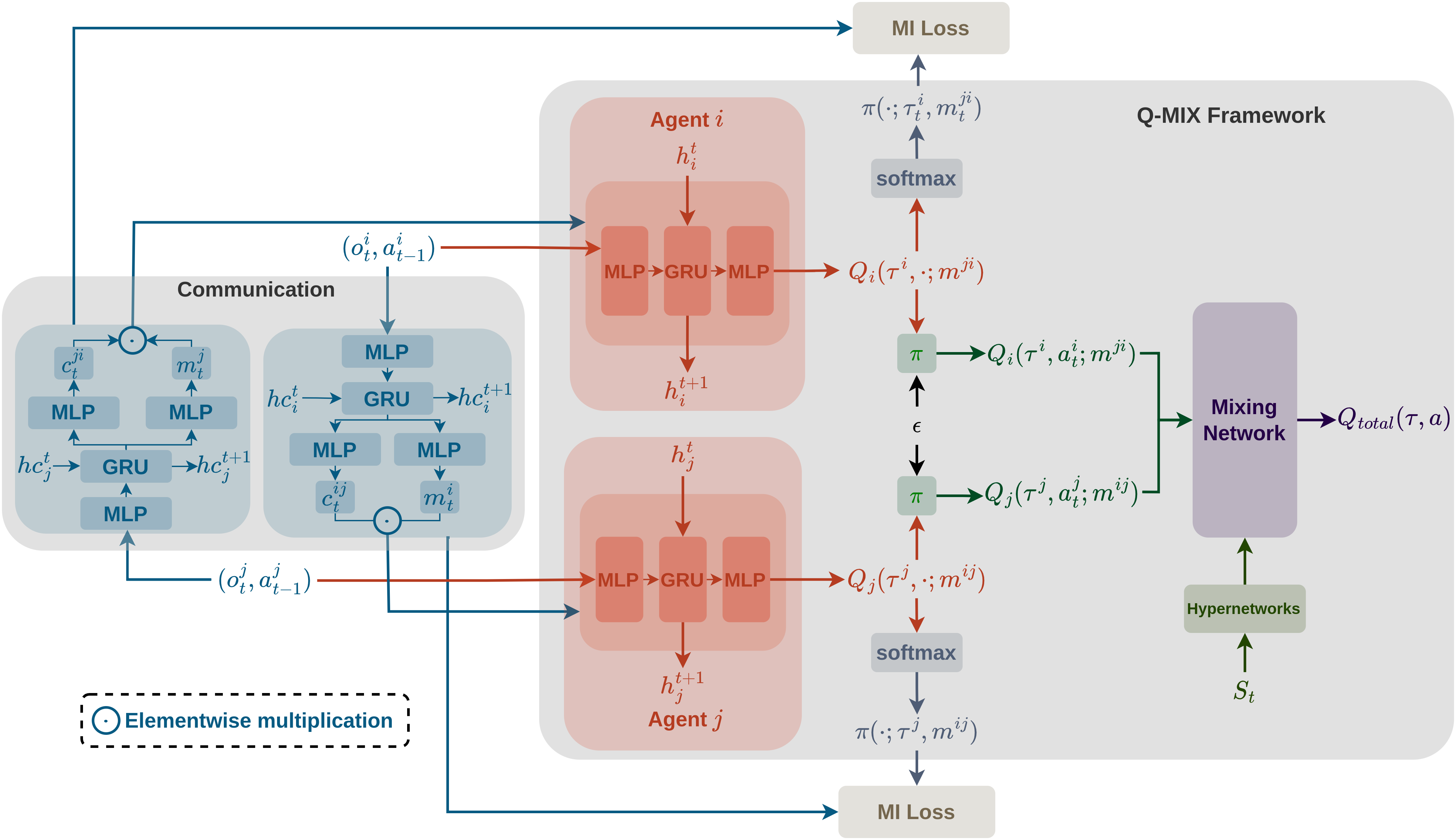}
    \caption{Architecture of QRC-TSC with two agents. Each agent uses a communication network (shown in the communication block) in addition to the agent network. The communication network takes the action-observation history $(o^{i}_{t}, a^{i}_{t-1})$ of the agent $i$ as input and outputs both the message $m_{t}^{ij}$ and a communication action $c_{t}^{ij}$ for the recipient $j$ at time $t$.}
    \label{fig:architecture}
\end{figure*}

In this section, we present a detailed design of QRC-TSC in the context of multi-agent Q-learning, Fig. \ref{fig:architecture}. We adopt the CTDE paradigm and use Q-MIX \cite{rashid2020monotonic} as a base learning framework. The training takes place in a centralized manner, assuming that the global state information is available. Each agent $i$ has access to an agent network with parameters shared across all agents. This approach has been shown to accelerate learning and enhance scalability in deep MARL settings. The agent network takes as inputs the action-observation history of the agent and the incoming messages from other agents to generate action-values. The agent uses its own action values to select an action during decentralized execution. Each agent also has a communication network that takes in the agent's action-observation history and generates the message vector $m_{ij}$ and a communication policy $c_{ij}$ for each available recipient agent $j \in \mathcal{N}(i)$. This can be seen in the communication module in Fig. \ref{fig:communication_concept}. The message is then gated $$\hat{m}_{ij} = (m \odot c)_{ij}$$ \footnote{$\odot$ represents elementwise multiplication} based on the communication action $c_{ij}$. The parameters of the communication network are also shared across agents. The mixing network combines the individual action-values of the agents $Q_{i}(\tau_{i}, a_{i}, \hat{m}_{ij}; \theta_{agent})$ to compute the join-action value function $Q_{total}$. The weights of the mixing network are generated by a set of hypernetworks conditioned on the state $s$. We use DIAL \cite{foerster2016learning} as the base communication framework and we improve on it in the following ways:
\begin{enumerate}
    \item We use variational inference to maximize the mutual information between the sent messages (including the communication action) and the recipient's action.
    \item We introduce an entropy regularization term for the communication policies, enabling controlled exploration in the communication action space.
    \item Communication policies are differentiable, allowing for end-to-end training.
\end{enumerate}

\subsection{Communication in QRC-TSC}

In our framework, each agent learns communication protocols through feedback from the recipient agents. Feedback is received in the form of gradients during backpropagation \cite{foerster2016learning,sukhbaatar2016learning}. Thus, the entire network architecture can be trained from a single objective function. Our goal in this work is to train agents to quickly and effectively learn communication protocols. Therefore, agents must learn the communication policy and send messages that reduce the uncertainty in the recipient's policy. To this end, we aim to maximize the mutual information between the sender's message and the recipient's policy, similar to NDQ \cite{wang2019learning}. This metric was previously proposed \cite{lowe2019pitfalls} as one of the key metrics to measure communication performance. Therefore, it makes sense to integrate such a metric into the objective function and explicitly maximize it.

First, we model outgoing messages as a joint distribution $p(m_{ij}, c_{ij})$ of the message generated $m_{ij}$ and its communication actions $c_{ij}$ by agent $i$ for agent $j$. 

\begin{align}
    p(m_{ij}, c_{ij} | \tau_{ij}) = p(m_{ij} | \tau_{ij})p(c_{ij} | \tau_{ij})
\end{align}

\cite{dupont2018learning}. Specifically, each agent $i$ generates a shared latent message distribution (a multivariate Gaussian) of size from which a message vector $\mathbf{m}_{i}$ is sampled and a discrete communication policy distribution (encoded as Bernoulli) which decides which bits of the messages are to be sent to agent $j$. 

This decision $c_{ij}$ is made independently for each agent $j \in \mathcal{N}(i)$ in the neighborhood. Similar to the approach proposed in \cite{jang2016categorical,maddison2016concrete}, we use Gumbel-sigmoid as a continuous approximation of the categorical variables. The Gumbel-max trick allows for differential sampling and does not suffer from high variance like the REINFORCE algorithm \cite{jang2016categorical}. Thus, our framework is end-to-end differentiable.

The communication action $c_{ij}$ acts as a mask over the messages during execution. To achieve this, we use differential relaxation of categorical/discrete variables \cite{jang2016categorical,maddison2016concrete}. Gumbel-sigmoid can be considered as a continuous relaxation of the Bernoulli distribution and can be written as

\begin{align}
    \sigma(\alpha_{l}) & = \text{sigmoid}((\alpha_{i} + g_{l} - g_{m}) / \mathcal{\lambda})
\end{align}

where $g_{l}$ and $g_{m}$ are samples from $Gumbel(0, 1)$ distribution and $\lambda$ is the temperature parameter.

Next, we discuss the objective function $J_{c}(\theta_{c})$ for learning communication. We maximize the mutual information between the sender's message and the recipient's policy.

\begin{align}
    \label{eq:mi_objective}
    I_{\theta_{c}}(\pi_{j}(\cdot | \tau_{j}); \hat{m}_{ij} | \tau_{j}, \hat{m}_{(-i)j}),
\end{align}

where $\pi_{j}(\cdot | \tau_{j}, \hat{m}_{j}^{in}) = \text{softmax}(Q(\cdot; \tau_{j}, \hat{m}_{j}^{in}))$ represents the policy of the agent conditioned on its action-observation history and incoming messages, $\hat{m}_{ij}$ is the resulting outgoing message from agent $i$ to agent $j$, and $\theta_{c}$ is the set of parameters of the communication network.

\begin{align}
    \label{eq:comm_mi_obj}
    J_{c}(\theta_{c}) = \sum_{j=1}^{n} [ \underbrace{I_{\theta_{c}}(\pi_{j}(\cdot | \tau_{j}); \hat{m}_{ij} | \tau_{j}, \hat{m}_{(-i)j})}_{(1)} - \underbrace{\beta I_{\theta{c}}(\hat{m}_{ij}; \tau_{i})}_{(2)} ],
\end{align}

where $\beta$ is the scaling factor that controls the tradeoff between the expressiveness and compressiveness of the messages. Since our objective is to maximize the mutual information it is sufficient to derive the objective as the lower bound for the term. The lower bound \cite{alemi2016deep,wang2019learning} for the mutual information objective, the first term in (\ref{eq:comm_mi_obj}) can be given as

\begin{align}
    \label{eq:comm_loss_1}
    \begin{split}
        & I_{\theta_{c}}(\pi_{j}(\cdot | \tau_{j}); \hat{m}_{ij} | \tau_{j}, \hat{m}_{(-i)j}) \\
        & \geq \mathbb{E}_{\boldsymbol{\tau} \sim \mathcal{D}, m^{in}_{ij}, c^{in}_{ij} \sim f_{c}(\tau; \theta_{c})}[ -\mathcal{CE}( \pi_{j}(\cdot; \tau_{j}, \hat{m}^{in}_{j} ) \| q_{\theta_{r}}( \cdot; \tau_{j}, \hat{m}^{in}_{j} ) ] \\
    \end{split}
\end{align}

where $\boldsymbol{\tau}$ is the joint local action-observation history of the agents sampled from the replay memory $\mathcal{D}$ and $\mathcal{CE}$ is the cross-entropy. The posterior estimates are given by $q_{\theta_{r}}( \cdot; \tau_{j}, \hat{m}^{in}_{j} ) = q_{\theta_{r}}( \cdot; \tau_{j}, (m \odot c)^{in}_{-(j)j} ) = q_{\theta_{r}}( \cdot; \tau_{j}, m^{in}_{-(j)j}, c^{in}_{-(j)j} )$ and parameters $\theta_{r}$ are shared across all the agents.

The second term, analogous to the variational bottleneck objective in \cite{alemi2016deep}, is the mutual information between the agent's action-observation history $\tau^{i}$ and the messages generated $m^{i}$.

\begin{align}
    \label{eq:comm_loss_2}
    \begin{split}
        \beta I_{\theta{c}}(\hat{m}_{ij}; \tau_{i}) & = \beta D_{KL}(p(m_{ij}, c_{ij} | \tau_{i}) \| q_{\theta_{r}}(m_{ij}, c_{ij} | \tau_{i})) \\
                                                    & = \underbrace{\beta_{m} D_{KL}(p(m_{ij} | \tau_{i}) \| q_{\theta_{r}}(m_{ij} | \tau_{i}))}_{(1)} \\
                                                    & \ + \underbrace{\beta_{c} D_{KL}(p(c_{ij} | \tau_{i}) \| q_{\theta_{r}}(c_{ij} | \tau_{i}))}_{(2)},
    \end{split}
\end{align}

The first term in (\ref{eq:comm_loss_2}) controls the tradeoff between maximizing the mutual information between the message $m_{ij}$ and agent $j$'s policy $\pi_{j}(\cdot; \tau_{j}, \hat{m}_{ij})$ and being compressive about the action-observation history $\tau_{i}$. The second term in (\ref{eq:comm_loss_2}) regularizes the communication policy. This encourages exploration of varied communication policies, which can be controlled by $\beta_{c}$.

Combining equations (\ref{eq:comm_loss_1}) and (\ref{eq:comm_loss_2}), we can write the loss function for the communication objective as:

\begin{align}
    \label{eq:comm_loss_total}
    \begin{split}
            & \mathcal{L}(\theta_{r}, \theta_{c}) = \\
            & \mathbb{E}_{\boldsymbol{\tau} \sim \mathcal{D}, m^{in}_{ij}, c^{in}_{ij} \sim f_{c}(\tau; \theta_{c})}[ \mathcal{CE}( \pi_{j}(\cdot; \tau_{j}, \hat{m}^{in}_{j} ) \| q_{\theta_{r}}( \cdot; \tau_{j}, \hat{m}^{in}_{j} ) ] \\
            & + \beta_{m} D_{KL}(p(m_{ij} | \tau_{i}) \| q_{\theta_{r}}(m_{ij} | \tau_{i})) \\
            & + \beta_{c} D_{KL}(p(c_{ij} | \tau_{i}) \| q_{\theta_{r}}(c_{ij} | \tau_{i}))
    \end{split}
\end{align}

Thus, the final loss function for training can be given as:

\begin{align}
    \label{eq:overall_loss}
    \mathcal{L}(\theta) = \mathcal{L}_{TD}(\theta) + \mathcal{L}_{C}(\theta_{r}, \theta_{c}),
\end{align}

where 
\begin{align}
    \label{eq:td_loss}
    \mathcal{L}_{TD} = [r + \gamma \max_{a'} Q_{total}(s', a'; \theta') - Q_{total}(s, a; \theta)]^{2}
\end{align} 
is the TD loss, $\theta^{-}$ is the set of parameters of the target network, $\theta$ is a set of parameters for all the networks combined and $\mathcal{L}_{C}(\theta_{r}, \theta_{c})$ is the total communication loss.

\begin{figure*}[!ht]
    \centering
    \subfloat[$4 \times 4$ grid network: Flow scenario 1]{\label{fig:4x4_scenario_flow_1}\includegraphics[width=.45\linewidth]{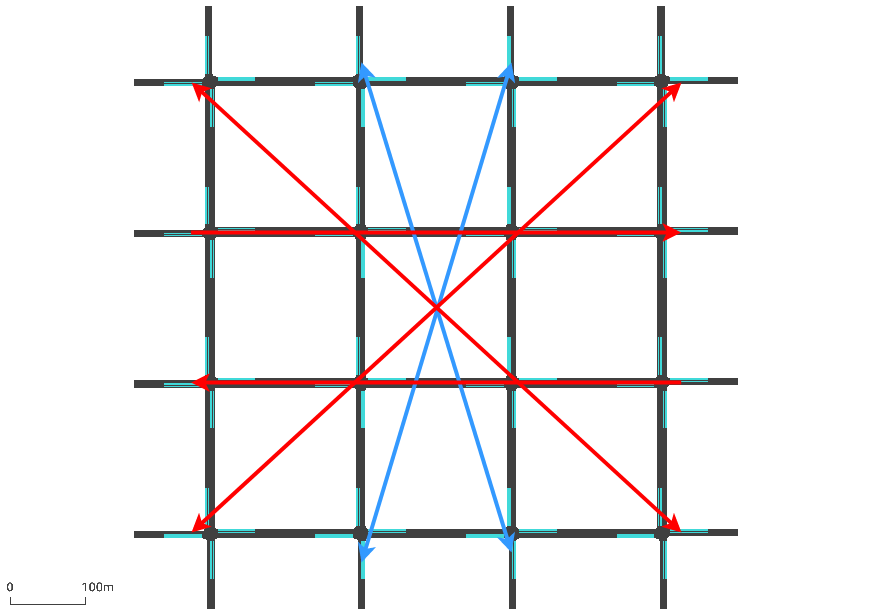}}
    \hfil
    \subfloat[$4 \times 4$ grid network: Flow scenario 2]{\label{fig:4x4_scenario_flow_2}\includegraphics[width=.45\linewidth]{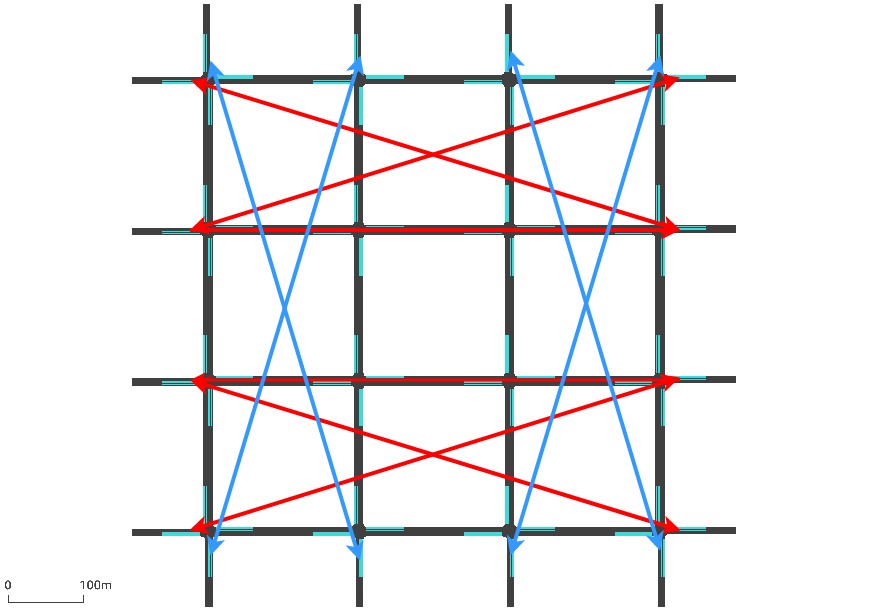}}
    \vskip\baselineskip
    \subfloat[Pasubio network]{\label{fig:pasubio_scenario_flow}\includegraphics[width=.45\linewidth]{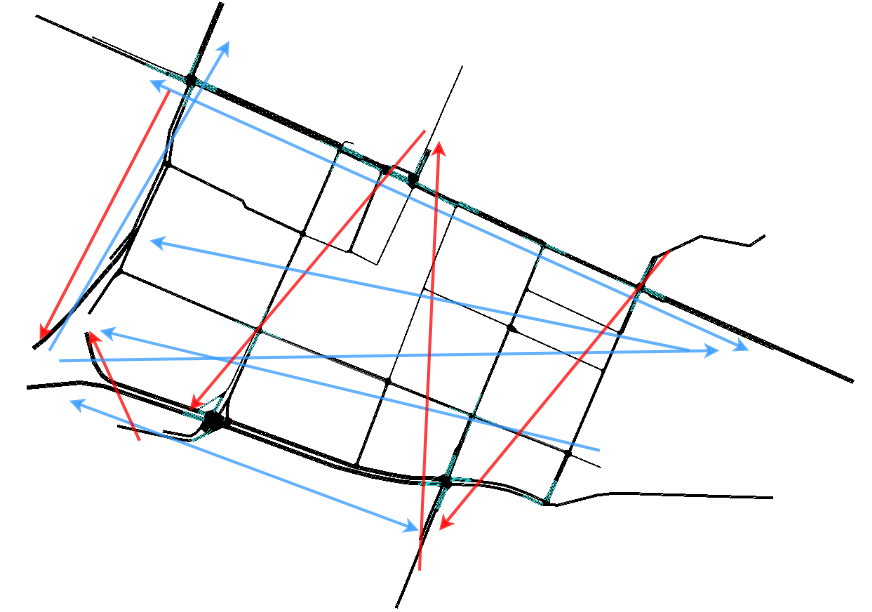}}
    \hfil
    \subfloat[Flow distribution]{\label{fig:flow_distribution}\includegraphics[width=.45\linewidth]{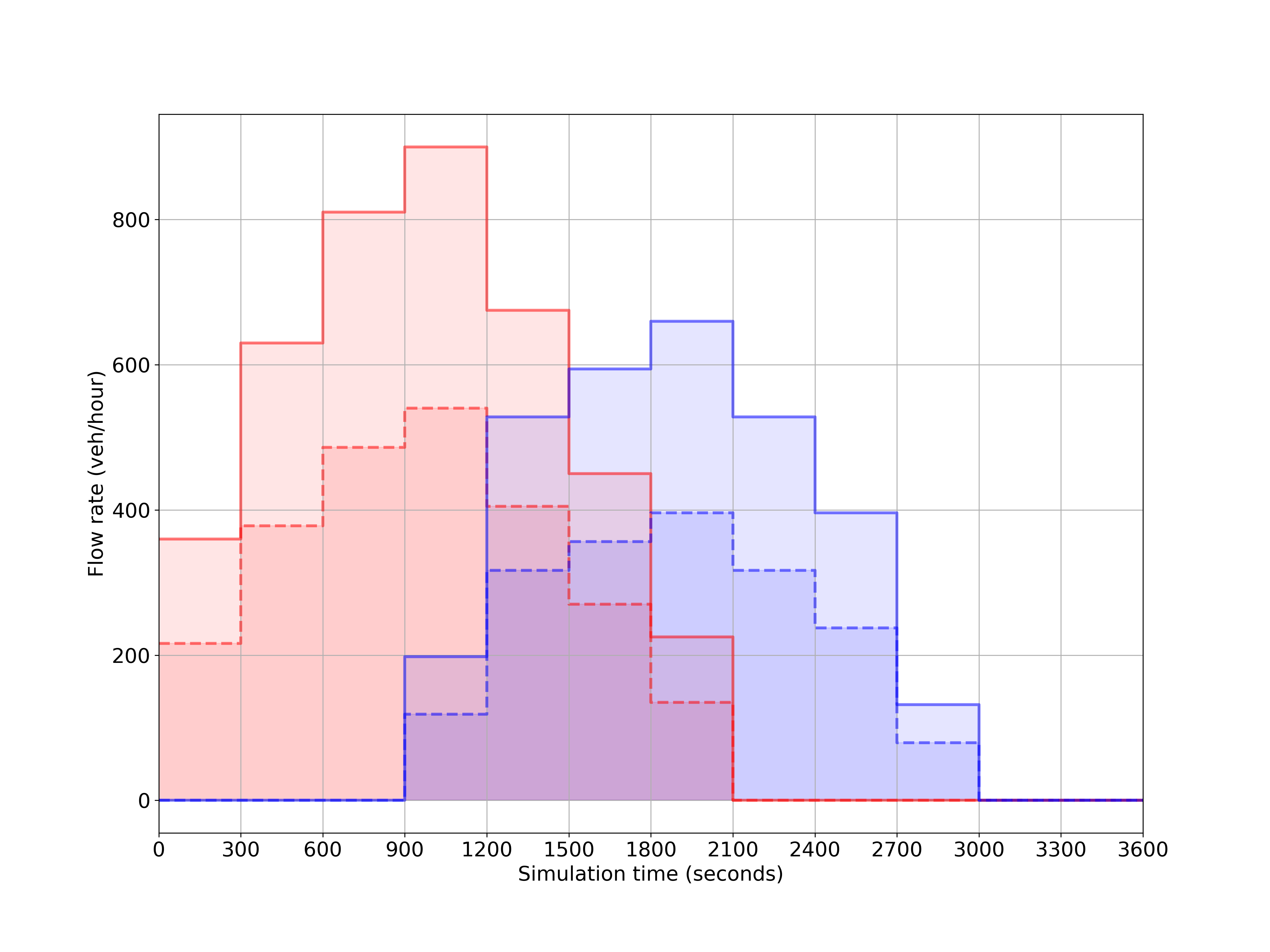}}
    \caption{(a) and (b) represent the flow scenarios for the $4 \times 4$ grid network. (c) shows the flow in Pasubio network and (d) shows the hourly flow distribution for both the networks. The dotted lines represent flow from opposite direction whenever bidirectional flows are simulated. The red and the blue lines represent the outer and inner network flow respectively.} 
    \label{fig:networks_and_flows}
\end{figure*}

%% file: experiments.tex
\section{Experiments}
\label{sec:experiments}

\subsection{Experimental Setup}

We built a synthetic $4 \times 4$ grid network and a real-world network of Pasubio, Bologna as proposed by Bieker et al. \cite{bieker2015traffic}. Trips are generated with origin-destination pairs of the fringe edges. For both the networks, we generated variable hourly traffic, similar to \cite{chu2019multi}, as shown in Fig. \ref{fig:flow_distribution}, where the solid lines represent the high flow rates and the dotted lines represent the low flow rates. Flow rates are varied in $5$-minute intervals within which the vehicles are inserted uniformly into the network with the specified flow rate. The peak flow rate is $900$ veh/hr. For convenience and representation purposes, we broke down the traffic flow into two types: (1) from east-west/west-east (red lines), which starts at the beginning of the hour and (2) north-south/south-north (blue lines), which starts after 15 minutes. Both flows last for 35 minutes. Flows from the opposite direction, represented by dotted lines in Fig. \ref{fig:flow_distribution}, are scaled down by a factor of $0.6$. Every hour, a random direction is selected as the opposite direction.

\begin{enumerate}

    \item \textbf{$4 \times 4$ grid network:}
        We built a two-lane synthetic $4 \times 4$ grid network of homogeneous agents. We simulated two traffic flow scenarios, one of which is selected randomly at the beginning of each simulation hour. For the first scenario, Fig. \ref{fig:4x4_scenario_flow_1}, we simulated high traffic on the external edges of the network, whereas in the second scenario Fig. \ref{fig:4x4_scenario_flow_2} internal edges of the network received a higher bulk of traffic flow. To induce a level of randomness in the traffic flow, a random direction was selected at the beginning of each simulation hour to have a high flow rate. Traffic flow settings in the synthetic version were not tethered to reality but were designed to test the robustness of the learning algorithm. The speed limit on all the lanes was around 14 m/s.

    \item \textbf{Pasubio network:}
        We used the real-world network of Pasubio, Bologna. The neighborhood has a hospital and includes common routes to the football stadium, and therefor is prone to congestion. The network has 7 traffic lights, some of which control multiple junctions. 3 traffic signals have 8 phases and the rest have 4, 10, 14, and 16 phases. The heterogeneity of the real-world network made it a more challenging environment than the synthetic network. We tried to replicate the traffic flow settings from \cite{bieker2015traffic}. The maximum allowable speed on each lane was set to 14 m/s.

\end{enumerate}

\begin{figure*}[!ht]
    \centering
    \subfloat[Average queue length in Pasubio network]{\label{fig:pasubio_mean_queues}\includegraphics[width=0.4\textwidth]{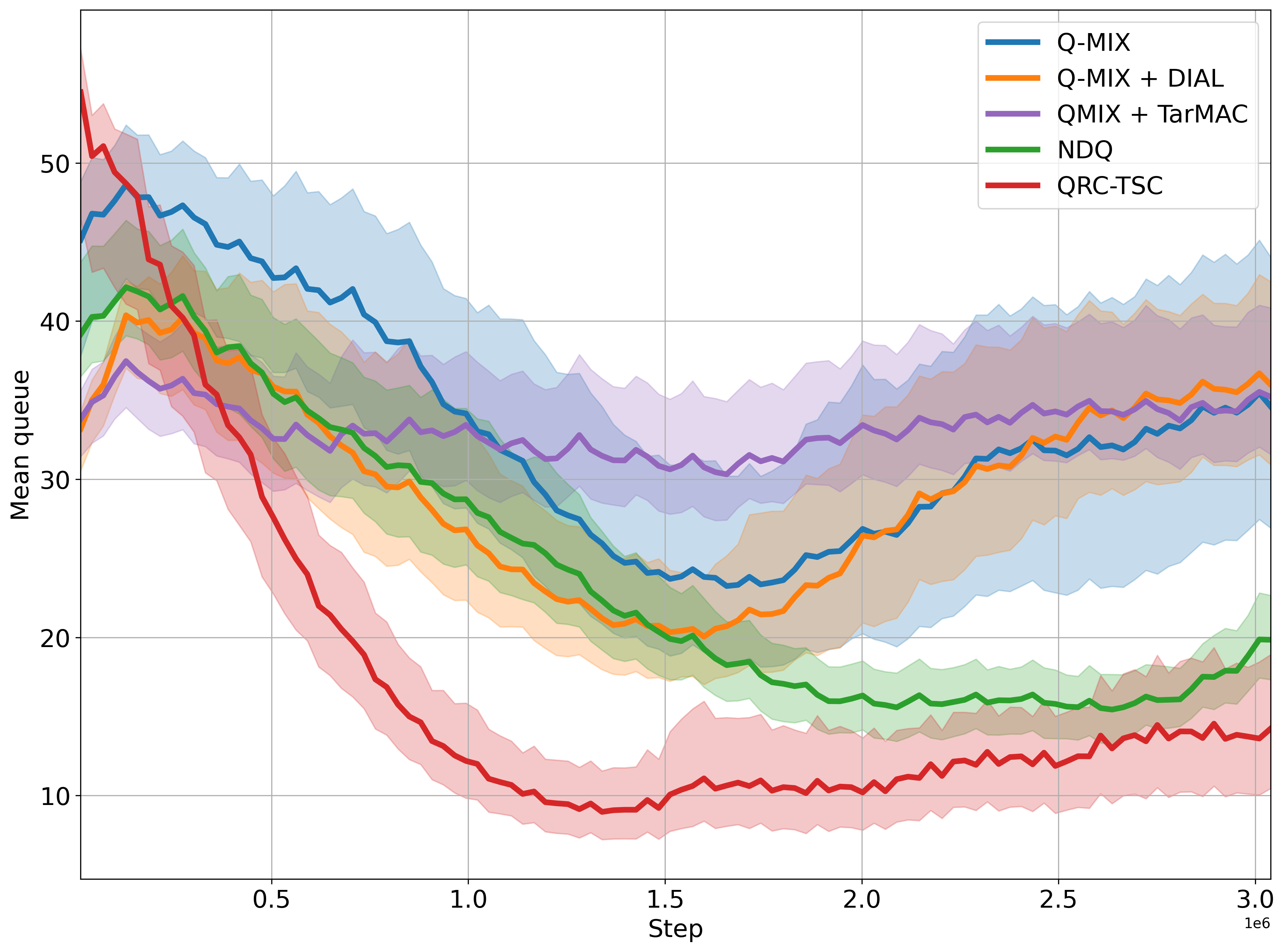}}
    \hfill
    \subfloat[Average queue length in $4 \times 4$ grid  network]{\label{fig:grid_mean_queues}\includegraphics[width=0.4\textwidth]{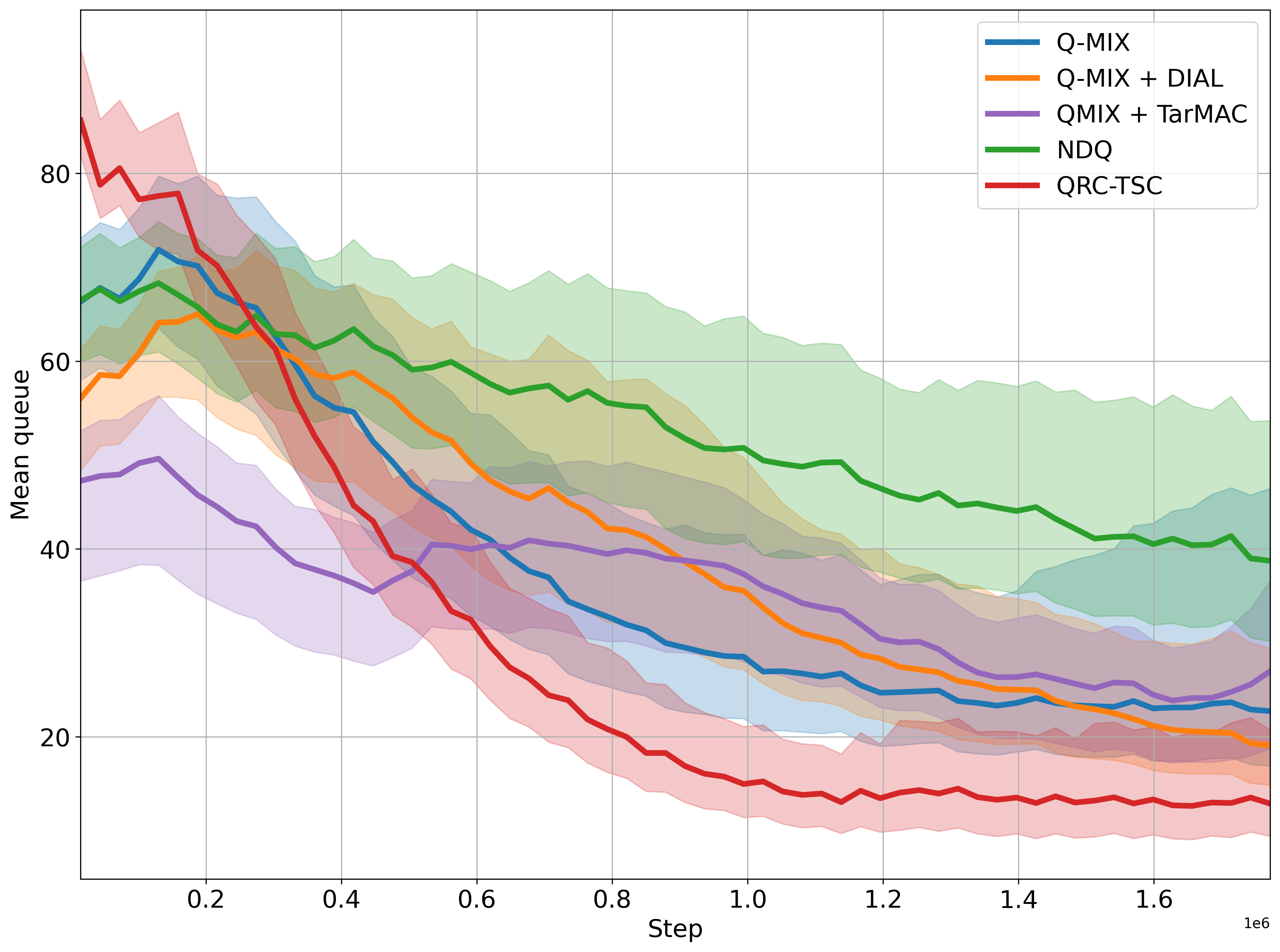}}
    \caption{The plot shows average queue length throughout training (lower the better). The x-axis represents simulation steps (in millions). The solid lines show mean over 5 runs and the shaded region represents $95\%$ CI.}
    \label{fig:results_mean_queues}
\end{figure*}

\begin{figure*}[!ht]
    \centering
    \includegraphics[width=\textwidth]{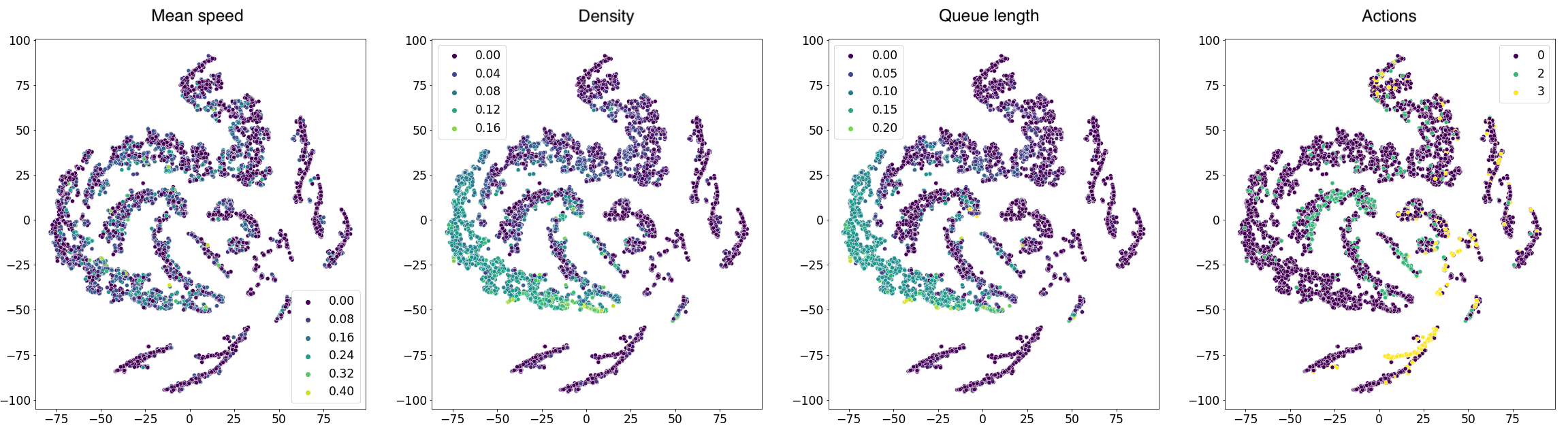}
    \caption{Message representation: The figure shows a t-SNE plot of the learned messages representations by an agent in the $4 \times 4$ grid network. The color scale in the first three plots, starting from left, represents a feature (averaged across all incoming lanes) of observations received by the agent. The color scale in the final plot represents the actions taken by the agent. We can see that the agent learns to embed messages in the latent space based on its inputs and action intentions.}
    \label{fig:message_representation}
\end{figure*}

Further, we adopt the metric average number of stops or queue length to measure the performance of the algorithms on the traffic network.

\subsection{Baselines}

In this work, we are interested in teaching the agents efficient communication policies. Specifically, our goal is to show that agents do not need to communicate all the time to be able to coordinate. Instead, agents can establish an optimal communication policy that tells the agent which parts of the message are worth sending and to which agent. To this end, we set Q-MIX \cite{rashid2020monotonic} as the baseline framework for learning the action-value function and DIAL \cite{foerster2016learning} as a baseline framework for communication. To make fair comparisons, we implemented DIAL by extending Q-MIX. We also compared our framework to NDQ \cite{wang2019learning}, a state-of-the-art method to learn communication, which uses thresholds to filter out unnecessary messages. Thus, all the methods we compared our framework to only differed in the type of communication mechanism: (i) Q-MIX can be seen as a base method without communication, (ii) Q-MIX + DIAL enables learning communication via a feedback mechanism, (iii) Q+MIX + TarMAC, adds attention mechanism to messages, and (iv) NDQ can be seen as an extension to Q-MIX + DIAL, which maximizes the mutual information between the sender's message and the recipient's policy. 

\subsection{Training Settings}

We trained all the algorithms on the grid network and Pasubio environment for 1.8 million and 3 million simulation steps, respectively. At the end of each episode, which lasted for 90 steps or 360 simulation seconds, we ran a training iteration. To evaluate the robustness of the algorithm, we ran 10 evaluation episodes with each agent selecting its actions greedily after every 200 training episodes.

\subsection{Results}

To ensure a fair comparison, we used Q-MIX as a baseline centralized training algorithm for all the algorithms based on communication. The learning curves of the algorithms are illustrated in Fig. \ref{fig:results_mean_queues}. The solid lines represent the hourly average queue length of an intersection for each scenario. Queue length, which represents the number of vehicles stopped in the incoming lanes of the traffic signal, is a key metric in evaluating the performance of a traffic signal network. Evaluations were conducted after every 200 training episodes, and the results were averaged over 15 independent runs. Additionally, we compare our algorithm to some traditional traffic signal control approaches (Fixed time \cite{roess2004traffic}, Self Organizing Traffic Lights (SOTL) \cite{gershenson2004self}, Max pressure \cite{varaiya2013max}). For the fixed time algorithm, the phase duration for green phases was set to 30 seconds.

In both network scenarios, QRC-TSC performed consistently better  than the other frameworks. While Q-MIX uses a centralized training mechanism to factorize the action-values, the agents operate in a completely decentralized way during execution. Purely decentralized policies can hinder the performance of systems, since traffic flow can be highly dynamic at times. On the other hand, in DIAL, the agents communicate all the time, which can decrease performance, as communication is often unnecessary and acts as additional noise. The performance of Q-MIX + DIAL, Q-MIX + TarMAC, and Q-MIX was relatively similar and significantly underperformed in the Pasubio scenario. The performance of NDQ and QRC-TSC was similar in the Pasubio network (Fig. \ref{fig:pasubio_mean_queues}), however, NDQ performed poorly in the grid network (Fig. \ref{fig:grid_mean_queues}). When considering average queue length, QRC-TSC consistently outperformed the other frameworks in both network scenarios and learned relatively stable policies, as can be seen in Fig. \ref{fig:results_mean_queues}.

\begin{table*}
    \caption{Performance results of various algorithms on $4 \times 4$ grid and Pasubio Network}
    \label{4x4_irregular_grid_table}
    \begin{tabularx}{\textwidth}{@{}l*{10}{C}c@{}}
        \toprule
        \multicolumn{9}{c}{$4 \times 4$ grid network} \\
        \midrule
        Metrics & {Q-MIX} & {DIAL + Q-MIX} & {TarMAC + Q-MIX} & {NDQ} & \textbf{QRC-TSC} & Fixed time & SOTL & Max pressure \\
        \midrule
        Mean queue length & 29.62 & 22.17 & 23.53 & 47.91 & \textbf{7.81} & 62.57 & 32.42 & 24.27 \\
        Mean wait time (s/veh) & 53.83 & 44.15 & 46.48 & 37.20 & \textbf{16.97} & 78.81 & 50.96 & 50.42 \\
        Mean speed (m/s) & 10.60 & 10.50 & 10.97 & 8.23 & \textbf{11.18} & 8.92 & 10.70 & 10.43 \\
        \% communication & 0 & 100 & 100 &  90.63 & \textbf{47.37} & 0 & 0 & 0 \\
        \midrule
        \multicolumn{9}{c}{Pasubio network} \\
        \midrule
        Mean queue length & 35.66 & 35.06 & 34.29 & 19.88 & \textbf{14.74} & 47.10 & 38.33 & 39.48 \\
        Mean wait time (s/veh) & 112.49 & 79.83 & 82.18 & 61.9 & \textbf{60.62} & 107.48 & 104.94 & 98.48 \\
        Mean speed (m/s) & 9.76 & 9.67 & 9.70 & 10.39 & \textbf{10.48} & 9.21 & 9.45 & 9.98 \\
        \% communication & 0 & 100 & 100 & 93.75 & \textbf{63.41} & 0 & 0 & 0 \\
        \bottomrule
    \end{tabularx}
\end{table*}

\subsection{Communication}

\subsubsection{Learned message representations}

Within our framework, each agent learns to generate messages conditioned on its action-observation history. Thus, messages can be interpreted as compressed representations of the agent's inputs and its action intentions. The message space is analogous to latent space in variational autoencoders, where each variable in the latent space is independent of the other. Thus, each bit in the message represents a unique information from the sender's action-observation history.  

Fig. \ref{fig:message_representation} shows an example t-SNE plot \cite{van2008visualizing} of message embeddings learned by our algorithm collected over 100 evaluation episodes. In the first three plots from left to right, the color gradients represent features of agents inputs averaged over the number of incoming lanes: mean speed ($\frac{1}{L} \sum_{l}s_{l}$), mean density ($\frac{1}{L} \sum_{l}n_{l}$), and mean queue length ($\frac{1}{L} \sum_{l}q_{l}$), respectively. These images show that the message distribution learned by the agents was correlated with its inputs, confirming that the agents learned to send meaningful information from their observations. The color labels in the fourth plot represent the actions taken by the agents, which indicates that the agents were able to effectively convey their action intentions through the message space. A key observation from this figure is that mean density and mean queue length are often correlated with each other, and hence the agent can eliminate information.

\subsubsection{Learned communication policies}

\begin{figure}[!ht]
    \centering
    \subfloat[Comparison of communication policies for $4 \times 4$ grid network]{\includegraphics[width=0.5\textwidth]{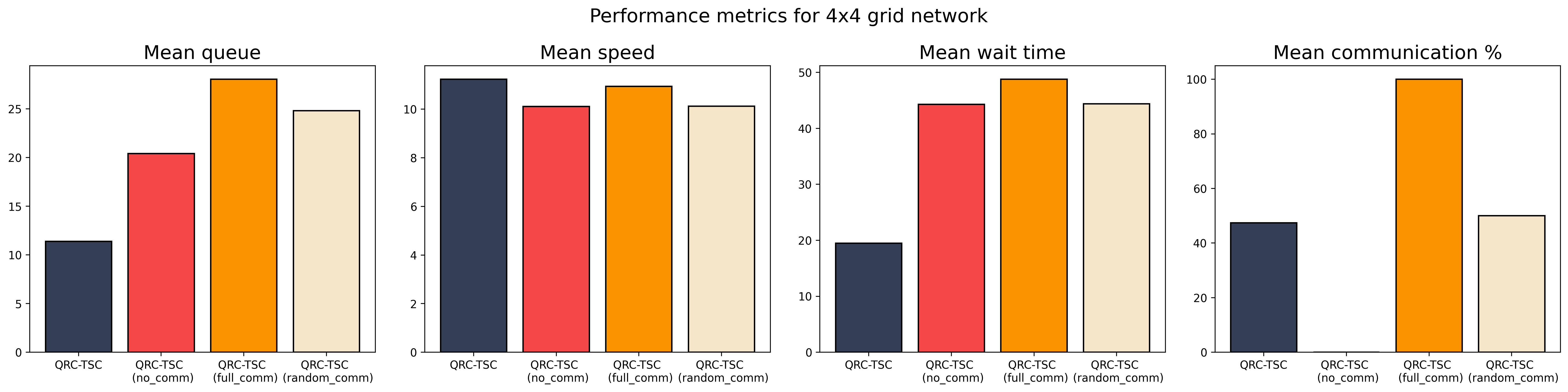}}
    \vfil
    \subfloat[Comparison of communication policies for Pasubio network]{\includegraphics[width=0.5\textwidth]{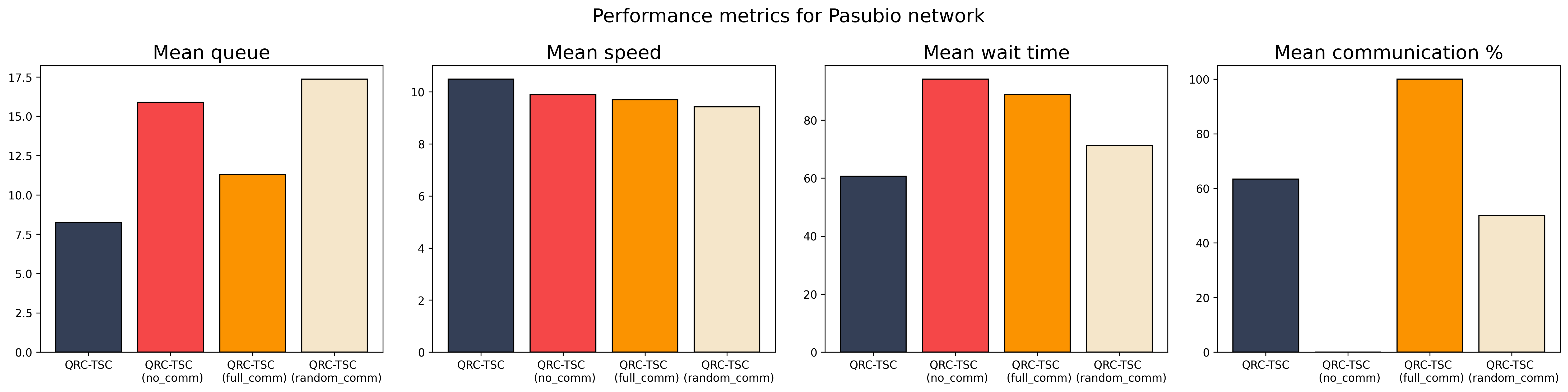}}
    \caption{Comparison of performance of communication policies averaged across 100 test episodes. QRC-TSC (in blue) represents the performance of the communication policies learned by our framework. }
    \label{fig:comm_comparison}
\end{figure}

In our framework, the agents are allowed to send 5 bit messages at each time step. Therefore, the communication policy can be seen as an action of selecting the bit of message for each recipient. This makes visualizing the communication policy for each agent in a reduced space almost infeasible. To evaluate the effectiveness of communication policy, we compared the communication policy learned by QRC-TSC with (1) random policy, (2) full communication, and (3) no communication. During the evaluation stage, we ran 3 additional independent tests where we manually altered the communication policies. Since this was done during the execution stage, we can be sure that altering the communication policies did not affect the training of QRC-TSC.

Fig. \ref{fig:comm_comparison} illustrates the performance of the communication policies learned by our framework. We selected a few key metrics (queue length, wait time, and mean speed) from traffic signal control theory to showcase the effectiveness of the learned policies. All metrics were averaged over 100 test episodes and across five runs. The performance of QRC-TSC (in blue) was the best across all metrics in both network scenarios. By choosing which bits to send, the agents were effectively able to balance the performance between no communication and full communication. It is interesting to note that the performance in the Pasubio network with full communication is the worst, which can also be seen in Fig. \ref{fig:pasubio_mean_queues}(a) where DIAL performs the worst among all the algorithms. This strongly indicates that constant communication can impede the performance of the system, likely caused due to redundancy in input information (from incoming messages). This leads us to conclude that the agents only need limited information about the action-observation history of the other agent to take optimal actions.

\subsection{Hyperparameters}

We based our framework on the PyMARL library \cite{samvelyan2019starcraft} and used the default parameters for all experiments. We experimented with different values for the message size and found that the message of length 5 performed the best. For the additional hyperparameters within the QRC-TSC framework, we conducted a coarse grid search to find the best set of hyperparameters. We set the value of both $\beta_{m}$ and $\beta_{c}$ to $10^{-5}$ across all environments. We tried linearly annealing the values of $\beta_{m}$ and $\beta_{c}$ over 50k iterations, but the overall performance change was negligible. We trained our models on an NVIDIA GeForce RTX 2080 using experience sampled from 8 parallel environments.

%% file: conclusion.tex
\section{Conclusion}
\label{sec:conclusion}

In this paper, we propose a novel communication mechanism enabling agents to effectively learn (i) \textit{which part of the message} is worth sending (ii) \textit{when} to send a message, and (iii) \textit{to whom} the message should be sent. This can be especially beneficial for problems where there exist constraints on communication (e.g. limited bandwidth). Further, our proposed framework is differentiable which allows for end-to-end training. The advantage of this framework is that the agents can act in a completely decentralized manner but exchange necessary bits of information to maintain coordination between agents. The framework is versatile and could be extended to a large number of applications. We tested our framework, QRC-TSC, on the real-world problem of traffic signal control by building two different traffic signal network scenarios (a synthetic and a real-world network). We compared QRC-TSC with several state-of-the-art frameworks involving communication, and demonstrate that it is able to maintain the least amount of congestion throughout the network while keeping the utility of the communication channel within $\sim 47-65$ percent. 

Some real-world problems have constraints, for example the cost of communication. Although this study did not address a constrained problem, we believe that QRC-TSC can be extended to include additional parameters, such as cost. One of the drawbacks QRC-TSC is that the maximum length of the message needs to be set a priori. One solution to this problem could be to allow for multiple communication passes. Future work will address how to establish the maximum message length.

%% file: acknowledgement.tex
\section{Acknowledgement}
\label{sec:acknowledgement}

This material is based upon work supported by the Northeastern University Tier 1 Award titled “Multi-Agent Reinforcement Learning Framework for Learning Coordination and Decision-Making”.

%% file: appendix.tex

\subsection{Algorithm for QRC-TSC}
\label{pseudocode}

\begin{algorithm}[H]
	\caption{Training procedure for QRC-TSC} 
	\begin{algorithmic}[1]
        \State Initialize the agent network with parameters $\theta$, the target network with parameters $\theta^{-}$, replay buffer $\mathcal{D}$ with capacity $N_{\mathcal{D}}$, and batch size $N_{B}$
        \For {each training episode $e$}
            \For {each episode}
                \State $t = 0 \ \text{and} \ h^{i}_{0} = \textbf{0}, \ hc^{i}_{0} = \textbf{0} \ \text{for each agent} \ i = \{ 1, \cdots, n \}$ 
                \While {$s_{t} \neq \text{terminal}$ \textbf{and} $t < T$}
                    \State $t = t + 1$
                    \State Obtain observation $o_{t} = \{ o^{1}_{t}, \cdots, o^{n}_{t} \}$ and global state $S_{t}$
                    \State Get message vector $\hat{m}^{i}_{t}$ and communication action $c^{i'}_{t}$ from agents \Comment{$\hat{m}^{i}_{t}, c^{i'}_{t} = \text{CNet}_{i}(o^{i}_{t}, m^{-i}_{t - 1}, hc^{i}_{t - 1}, a^{i}_{t - 1}; \theta^{i}_{c})$}
                    \State Set outgoing messages as: $m^{i'}_{t} = \hat{m}^{i}_{t} \odot c^{i'}_{t}$
                    \State Select action $a^{i}_{t}$ according to $\epsilon$-greedy policy w.r.t. agent $i$'s decentralized action value $Q(o^{i}_{t}, m^{-i}_{t - 1}, h^{i}_{t - 1}, a^{i}_{t - 1}; \theta^{i})$
                    \State Execute joint action $a_{t} = \{ a^{1}_{t}, \cdots, a^{n}_{t} \}$ in the environment
                    \State Obtain the global reward $r_{t + 1}$, next observation $o^{i}_{t + 1}$ for each agent $i$ and next global state $s_{t + 1}$
                \EndWhile
                \State Store the episode in the buffer $\mathcal{D}$ such that the oldest episode is replaced if $| \mathcal{D} | \geq N_{\mathcal{D}}$
            \EndFor
            \State Sample a batch of $N_{B}$ episodes $\sim \text{Uniform}(\mathcal{D})$
            \State Calculate the communication loss $\mathcal{L}_{C}(\theta_{r}, \theta_{c})$ according to (\ref{eq:comm_loss_total}) and TD loss $\mathcal{L}_{TD}(\theta)$ as in (\ref{eq:td_loss}) and set total loss as in (\ref{eq:overall_loss})
            \State Update $\theta$ by minimizing the total loss $\mathcal{L}(\theta)$
            \State Replace target parameters $\theta^{-} \leftarrow \theta$ every $K$ episodes
        \EndFor
	\end{algorithmic} 
\end{algorithm}

\subsection{Variational Bound on Mutual Information}

The posterior for the mutual information objective based on information bottleneck \cite{alemi2016deep} can be written as

\begin{align*}
    & I_{\theta_{c}}(\pi_{j}(\cdot | \tau_{j}); \hat{m}_{ij} | \tau_{j}) \\
    & = \int p(\tau_{j}) \pi_{j}(\cdot | \tau_{j}) p(\hat{m}_{ij} | \tau_{j}, \hat{m}_{(-i)j}) \log \pi_{j}(\cdot | \tau_{j}, \hat{m}_{ij}) d\tau_{j} d\pi_{j} d\hat{m}_{ij} \\
\end{align*}
Next, $q_{\theta_{r}}( \cdot; \tau_{j}, \hat{m}^{in}_{j} )$ can be written as variational approximation to $\pi_{j}(\cdot | \tau_{j}, \hat{m}_{ij})$ and since $D_{KL}( \pi_{j}(\cdot; \tau_{j}, \hat{m}^{in}_{j} ) \| q_{\theta_{r}}( \cdot; \tau_{j}, \hat{m}^{in}_{j} ) \geq 0$, we obtain the upper bound for mutual information term. 
\begin{align*}
     & \geq \int p(\tau_{j}) \pi(\cdot | \tau_{j}) p(\hat{m}_{ij} | \tau_{j}, \hat{m}_{(-i)j}) \log q_{\theta_{r}}(\cdot | \tau_{j}, \hat{m}_{ij}) d\tau_{j} d\pi_{j} d\hat{m}_{ij} \\
\end{align*}
We approximate $p(\tau_{j}, a_{j}) = p(\tau_{j}) \pi(\cdot | \tau_{j})$ using Monte Carlo sampling.
\begin{align*}
    & \geq \mathbb{E}_{\boldsymbol{\tau} \sim \mathcal{D}, \hat{m}^{in}_{j} \sim f_{c}(\tau; \theta_{c})} [ \int \pi(\cdot | \tau_{j}) \log q_{\theta_{r}}(\cdot | \tau_{j}, \hat{m}_{j}^{in}) d\pi_{j} ] + \mathcal{H}(\cdot | \tau_{j}, \hat{m}_{(-i)j}) \\
    & \geq \mathbb{E}_{\boldsymbol{\tau} \sim \mathcal{D}, m^{in}_{ij}, c^{in}_{ij} \sim f_{c}(\tau; \theta_{c})}[ -\mathcal{CE}( \pi_{j}(\cdot; \tau_{j}, \hat{m}^{in}_{j} ) \| q_{\theta_{r}}( \cdot; \tau_{j}, \hat{m}^{in}_{j} ) ], \\
\end{align*}
where entropy term $\mathcal{H}(\cdot | \tau_{j}, \hat{m}_{(-i)j})$ is independent of optimization.


\subsection{Communication loss function for joint distributions}
\label{communication_loss}

\begin{align*}
            & J_{c}[\hat{m}_{i} | \tau_{j}, \hat{m}_{(-i)j}] \\
            & \geq \mathbb{E}_{\boldsymbol{\tau} \sim \mathcal{D}}[-\mathcal{CE}[p(a_{j} | \boldsymbol{\tau}) || q_{\theta_{r}}(a_{j} | \boldsymbol{\tau}, \hat{m}_{j}^{in})]] - E_{m_{j}^{in}, c_{j}^{in} \sim f_{c}(\boldsymbol{\tau}; \theta_{c})} [  D_{KL}(p(\hat{m}_{ij} | \boldsymbol{\tau}_{i}) || r(\hat{m}_{ij}))] \\
            & \geq \mathbb{E}_{\boldsymbol{\tau} \sim \mathcal{D}}[-\mathcal{CE}[p(a_{j} | \boldsymbol{\tau}) || q_{\theta_{r}}(a_{j} | \boldsymbol{\tau}, (m \odot c)_{j}^{in})]] - E_{m_{j}^{in}, c_{j}^{in} \sim f_{c}(\boldsymbol{\tau}; \theta_{c})} \left[ D_{KL}\left( \log \frac{p(m_{ij}, c_{ij} | \boldsymbol{\tau}_{i})}{r(m_{ij}, c_{ij})} \right) \right] \\
            & \geq \mathbb{E}_{\boldsymbol{\tau} \sim \mathcal{D}}[-\mathcal{CE}[p(a_{j} | \boldsymbol{\tau}) || q_{\theta_{r}}(a_{j} | \boldsymbol{\tau}, (m \odot c)_{j}^{in})]] - E_{m_{j}^{in}, c_{j}^{in} \sim f_{c}(\boldsymbol{\tau}; \theta_{c})} \left[   D_{KL} \left( \log \frac{p(m_{ij} | \boldsymbol{\tau}_{i})p(c_{ij} | \boldsymbol{\tau}_{i})}{r(m_{ij})r(c_{ij})} \right) \right] \\
            & \geq \mathbb{E}_{\boldsymbol{\tau} \sim \mathcal{D}}[-\mathcal{CE}[p(a_{j} | \boldsymbol{\tau}) || q_{\theta_{r}}(a_{j} | \boldsymbol{\tau}, (m \odot c)_{j}^{in})]] \\ 
            & \begin{aligned}[t]
            & \quad - E_{m_{j}^{in} \sim f_{c}(\boldsymbol{\tau}; \theta_{c})} \left[ D_{KL} \left( \log \frac{p(m_{ij} | \boldsymbol{\tau}_{i})}{r(m_{ij})} \right) \right] - E_{c_{j}^{in} \sim f_{c}(\boldsymbol{\tau}; \theta_{c})} \left[ D_{KL} \left( \log \frac{p(c_{ij} | \boldsymbol{\tau}_{i})}{r(c_{ij})} \right) \right] \\ \end{aligned}
\end{align*}

\subsection{Gumbel-Sigmoid for discrete communication variable}

We consider the communication action $c_{ijk}$ as a Bernoulli random variable. We drop the subscripts $ijk$ for the ease of notation. Let $c \sim Bernoulli(\alpha)$ be the communication action, where $\alpha \in (0, \infty)$ is the location parameter. We can write the Gumbel-softmax \cite{gumbel1954some, jang2016categorical} function as:
\begin{align*}
    c_{l} = \frac{\exp((\log \alpha_{l} + g_{l}) / \lambda)}{\sum_{l} \exp((\log \alpha_{l} + g_{l}) / \lambda)} \\
\end{align*}
where $\lambda \in (0, \infty)$ is the temparature parameter and $l$ is the dimension over the softmax vector. And we can rewrite the softmax function for two variables $\alpha_{l}$ and $0$ as:
\begin{align*}
    \begin{split}
        \sigma(\alpha_{l})
                    & = \frac{\exp((\log \alpha_{l} + g_{l}) / \lambda)}{\exp((\log \alpha_{l} + g_{l}) / \lambda) + \exp(g_{m} / \lambda)} \\
                    & = \frac{1}{1 + (\exp(g_{m}/\lambda) / \exp((\alpha_{i} + g_{l})/ \lambda))} \\
                    & = \frac{1}{1 + \exp(-(\log \alpha_{i} + g_{l} - g_{m}) / \lambda)} \\
                    & = sigmoid((\log \alpha_{i} + g_{l} - g_{m}) / \lambda)
    \end{split}
\end{align*}
The difference between two Gumbel distributions $g_{l} - g_{m}$ is given as Logistic distribution and can be sampled as $ \log U - \log(1 - U)$, where $U \sim Uniform(0, 1)$ \cite{maddison2016concrete}. We set the value of $\lambda$ to 0.67.